\documentclass[12pt]{article}
\usepackage{graphicx}
\usepackage{lscape}
\usepackage{citesort}
\usepackage{amssymb}
\usepackage{appendix}
\usepackage{multirow}

\begin{document}

\def\Tr{\mbox{Tr}}
\def\figt#1#2#3{
        \begin{figure}
        $\left. \right.$
        \vspace*{-2cm}
        \begin{center}
        \includegraphics[width=10cm]{#1}
        \end{center}
        \vspace*{-0.2cm}
        \caption{#3}
        \label{#2}
        \end{figure}
	}
	
\def\figb#1#2#3{
        \begin{figure}
        $\left. \right.$
        \vspace*{-1cm}
        \begin{center}
        \includegraphics[width=10cm]{#1}
        \end{center}
        \vspace*{-0.2cm}
        \caption{#3}
        \label{#2}
        \end{figure}
                }

\def\ds{\displaystyle}
\def\beq{\begin{equation}}
\def\eeq{\end{equation}}
\def\bea{\begin{eqnarray}}
\def\eea{\end{eqnarray}}
\def\beeq{\begin{eqnarray}}
\def\eeeq{\end{eqnarray}}
\def\ve{\vert}
\def\vel{\left|}
\def\ver{\right|}
\def\nnb{\nonumber}
\def\ga{\left(}
\def\dr{\right)}
\def\aga{\left\{}
\def\adr{\right\}}
\def\lla{\left<}
\def\rra{\right>}
\def\rar{\rightarrow}
\def\lrar{\leftrightarrow}  
\def\nnb{\nonumber}
\def\la{\langle}
\def\ra{\rangle}
\def\ba{\begin{array}}
\def\ea{\end{array}}
\def\tr{\mbox{Tr}}
\def\ssp{{\Sigma^{*+}}}
\def\sso{{\Sigma^{*0}}}
\def\ssm{{\Sigma^{*-}}}
\def\xis0{{\Xi^{*0}}}
\def\xism{{\Xi^{*-}}}
\def\qs{\la \bar s s \ra}
\def\qu{\la \bar u u \ra}
\def\qd{\la \bar d d \ra}
\def\qq{\la \bar q q \ra}
\def\GG{\langle g_s^2 G^2 \rangle}
\def\q{\gamma_5 \not\!q}
\def\x{\gamma_5 \not\!x}
\def\g5{\gamma_5}
\def\sb{S_Q^{cf}}
\def\sd{S_d^{be}}
\def\su{S_u^{ad}}
\def\sbp{{S}_Q^{'cf}}
\def\sdp{{S}_d^{'be}}
\def\sup{{S}_u^{'ad}}
\def\ssp{{S}_s^{'??}}

\def\sig{\sigma_{\mu \nu} \gamma_5 p^\mu q^\nu}
\def\fo{f_0(\frac{s_0}{M^2})}
\def\ffi{f_1(\frac{s_0}{M^2})}
\def\fii{f_2(\frac{s_0}{M^2})}
\def\O{{\cal O}}
\def\sl{{\Sigma^0 \Lambda}}
\def\es{\!\!\! &=& \!\!\!}
\def\ap{\!\!\! &\approx& \!\!\!}
\def\ar{&+& \!\!\!}
\def\ek{&-& \!\!\!}
\def\kek{\!\!\!&-& \!\!\!}
\def\cp{&\times& \!\!\!}
\def\se{\!\!\! &\simeq& \!\!\!}
\def\eqv{&\equiv& \!\!\!}
\def\kpm{&\pm& \!\!\!}
\def\kmp{&\mp& \!\!\!}
\def\mcdot{\!\cdot\!}
\def\erar{&\rightarrow&}


\def\simlt{\stackrel{<}{{}_\sim}}
\def\simgt{\stackrel{>}{{}_\sim}}


\title{
         {\Large
                 {\bf
Electromagnetic transitions among octet and decuplet baryons in QCD
                 }
         }
      }

\author{\vspace{1cm}\\
{\small T. M. Aliev \thanks {e-mail:
taliev@metu.edu.tr}~\footnote{permanent address: Institute of
Physics, Baku, Azerbaijan}\,\,, Y. \"{O}ktem \thanks {e-mail:
sgyks@istanbul.edu.tr}\,\,, M. Savc{\i} \thanks
{e-mail: savci@metu.edu.tr}} \\
{\small Physics Department, Middle East Technical University,
06531 Ankara, Turkey }\\
{\small$^\ddag$ Physics Department,  Faculty of Sciences,
\.{I}stanbul University,} \\
{\small Vezneciler,  34134 \.{I}stanbul, Turkey}}

\date{}

\begin{titlepage}
\maketitle
\thispagestyle{empty}

\begin{abstract}

The magnetic dipole $G_M(Q^2)$, electric quadrupole $G_E(Q^2)$, and Coulomb
quadrupole $G_C(Q^2)$ form factors, describing the spin--3/2 to spin--1/2
electromagnetic transitions, are investigated within the light cone QCD sum
rules. The $Q^2$ dependence of these form factors, as well as ratios of
electric quadrupole and Coulomb quadrupole form factors to the magnetic
dipole form factors are studied. We also compare our results on the magnetic
dipole form factor with the prediction of the covariant spectator quark
model.

\end{abstract}

~~~PACS numbers: 11.55.Hx, 13.30.Ce, 13.40.Gp, 14.20.Jn

\end{titlepage}

\section{Introduction}

According to the SU(3) classification of the baryons, spin--1/2 and
spin--3/2 baryons belong to the octet and decuplet representations,
respectively. The electromagnetic properties of octet and decuplet baryons,
as well as octet to decuplet transitions are characterized by their
electromagnetic form factors. As is well known, form factors carry essential
information about the internal structure of baryons, i.e., about their
charge and current distribution. The octet to decuplet electromagnetic
transition is described with the help of the magnetic dipole (M1), electric
quadrupole (E2) and Coulomb quadrupole (C2) form factors, which follow from
the spin--parity selection rule.

At present, rich experimental data have been accumulated on
electromagnetic and
$\gamma^\ast N \to \Delta$ transition form factors (see \cite{Rodt01} and
references therein), while the data for other possible $\gamma^\ast~octet
\to decuplet$ transitions are practically absent (very limited data can be found in
\cite{Rodt02,Rodt03}). The octet to decuplet electromagnetic transitions are
studied within different theoretical approaches, such as the quark model
\cite{Rodt04,Rodt05,Rodt06,Rodt07}, QCD sum rules \cite{Rodt08,Rodt09},
and lattice QCD \cite{Rodt10}. In many theoretical works, the octet to decuplet
electromagnetic transitions have been studied at $Q^2=0$, but only in a few
works have the form factors of these transitions been studied. These
form factors were studied in the framework of the covariant quark
model in \cite{Rodt11} and in lattice QCD in \cite{Rodt12}, for the
$\gamma^\ast N \to \Delta$ transition only, and were investigated in light cone QCD
sum rules in \cite{Rodt13}, respectively.

In the present work we study the octet to decuplet transition form factors
within the light cone QCD sum rules. Note that these form factors are calculated
within the same approach at $Q^2=0$ in \cite{Rodt09}. The plan of the work
is as follows. In the
following section we derive the sum rules for the form factors responsible
for the octet to decuplet electromagnetic transition, whose numerical
analysis is performed in Sec. 3. This section also contains a comparison of
our results with other approaches and conclusions.

\section{Sum rules for the octet to decuplet electromagnetic transition form
factors}

In the present section we derive sum rules for the octet to decuplet
electromagnetic transition form factors. For this purpose we consider the
following correlator function,
\bea
\label{eodt01}
\Pi_{\mu\nu}(p,q) = i \int d^4x e^{iqx} \lla 0 \vel T \Big\{\eta_\mu(0)
j_\nu^{el}(x) \Big\} \ver {\cal O}(p) \rra~,
\eea
where $\eta_\mu$ is the interpolating current for the relevant decuplet baryon;
${\cal O}(p)$ represents an octet baryon with momentum $p$; $j_\nu^{el} = e_u \bar{u}
\gamma_\nu u + e_d \bar{d} \gamma_\nu d + e_s \bar{s} \gamma_\nu s$
is the electromagnetic current; $q$ is its four--momentum; and $e_u$,
$e_d$, and $e_s$ are the charges of $u$, $d$ and $s$ quarks, respectively.
The interpolating current of decuplet baryons can be written as,
\bea
\label{nolabel}
\eta_\mu = N \varepsilon^{abc} \Big\{ \left( q_1^{aT} C \gamma_\mu q_2^b
\right) q_3^c + \left( q_2^{aT} C \gamma_\mu q_3^b \right) q_1^c +
\left( q_3^{aT} C \gamma_\mu q_1^b \right) q_2^c \Big\}~,\nnb
\eea
where $a$, $b$, and $c$ are the color indices and $C$ is the charge
conjugation operator. The quark content $q_1$, $q_2$, and $q_3$
and the normalization factor $N$ of each decuplet baryon are given in Table
1.


\begin{table}[h]

\renewcommand{\arraystretch}{1.3}
\addtolength{\arraycolsep}{-0.5pt}
\small
$$
\begin{array}{|l|c|c|c|c|}
\hline \hline
                  &     N        & q_1 & q_2 & q_3 \\  \hline
 \Sigma^{\ast +}  & \sqrt{1/3}   &  u  &  u  &  s  \\
 \Sigma^{\ast 0}  & \sqrt{2/3}   &  u  &  d  &  s  \\
 \Sigma^{\ast -}  & \sqrt{1/3}   &  d  &  d  &  s  \\
 \Xi^{\ast 0}     & \sqrt{1/3}   &  s  &  s  &  u  \\
 \Xi^{\ast -}     & \sqrt{1/3}   &  s  &  s  &  d  \\
\hline \hline  
\end{array}
$$
\caption{Quark content and the value of normalization factor $N$ in
interpolating current of the decuplet baryons}
\renewcommand{\arraystretch}{1}
\addtolength{\arraycolsep}{-1.0pt}
\end{table}


The form factors which describe the octet to decuplet electromagnetic
transition are defined by the matrix element of the $j_\nu^{el}$ sandwiched
between the members of the decuplet and octet baryon states with momenta
$p^\prime$ and $p$, respectively. By virtue of Lorentz invariance and
current conservation, this matrix element can be written in terms of three
form factors as follows \cite{Rodt14}:

\bea
\label{eodt02}
\lla {\cal D} (p^{\prime}) \vel j_\nu^{el} \ver {\cal O} (p) \rra \es 
\bar{u}_\alpha(p^{\prime}) \Bigg\{ G_1(Q^2) \Big( -q_\alpha \gamma_\nu +
\rlap/{q}g_{\alpha\nu}\Big) + {G_2(Q^2)\over 2} \Big[ -q_\alpha {\cal P}_\nu
+ (q\mcdot {\cal P}) g_{\alpha\nu} \Big] \nnb \\
\ar  G_3(Q^2) \Big[ q_\alpha q_\nu - q^2
g_{\alpha\nu} \Big] \Bigg\} \gamma_5 u(p)~,
\eea
where ${\cal P} = {1\over 2} (p^{\prime} + p) = {1\over 2} (2 p - q)$, and
$u_\alpha(p^{\prime})$ is the spin--3/2 Rarita--Schwinger spinor.
From an experimental point of view, the so--called multipole form factors are
more suitable and we shall use this set of form factors 
in further analysis. The relations among the form
factors $G_1$, $G_2$, $G_3$ and the multipole form factors, namely,
magnetic dipole $G_M(Q^2)$, electric
quadrupole $G_E(Q^2)$, and Coulomb quadrupole $G_C(Q^2)$ form factors, are
given as \cite{Rodt14}  
\bea
\label{eodt03}
G_M(Q^2) \es {m_{\cal D} \over 3 (m_{\cal D} + m_{\cal O})}
\Bigg\{\Big[(3 m_{\cal D} + m_{\cal O}) (m_{\cal D} + m_{\cal O}) +
Q^2\Big] {G_1(Q^2) \over m_{\cal D}} \nnb \\
\ar (m_{\cal D}^2 - m_{\cal O}^2) G_2(Q^2) - 2 Q^2 G_3(Q^2)\Bigg\}~, \nnb \\ 
G_E(Q^2) \es {m_{\cal D} \over 3 (m_{\cal D} + m_{\cal O})}
\Bigg\{(m_{\cal D}^2 - m_{\cal O}^2 - Q^2){G_1(Q^2)\over m_{\cal D}} +
(m_{\cal D}^2 - m_{\cal O}^2) G_2(Q^2) - 2 Q^2 G_3(Q^2) \Bigg\}~, \nnb \\
G_C(Q^2) \es {2 m_{\cal D} \over 3 (m_{\cal D} + m_{\cal O})}
\Bigg\{2 m_{\cal D} G_1(Q^2) +
(3 m_{\cal D}^2 + m_{\cal O}^2 + Q^2) {G_2(Q^2)\over 2} \nnb \\
\ar (m_{\cal D}^2 - m_{\cal O}^2 - Q^2) G_3(Q^2) \Bigg\}~.
\eea
In further analysis we shall also study the dependence of the ratios of the
electric $G_E(Q^2)$ and Coulomb $G_C(Q^2)$ quadrupole form factors to the
dipole form factor $G_M(Q^2)$. Note that these ratios are measured in experiments
for the $\gamma^\ast N \to \Delta$ transitions which are defined as
\bea
\label{eodt04}
R_{EM}(Q^2) \es - {G_E(Q^2) \over G_M(Q^2)}~, \nnb \\
R_{SM}(Q^2) \es - {G_C(Q^2) \over 2 m_{\cal D} G_M(Q^2)}
\sqrt{Q^2 + {(m_{\cal D}^2 - m_{\cal O}^2 - Q^2)^2 \over
m_{\cal D}^2}}~.
\eea

In order to derive the sum rules for the octet to decuplet transition form
factors we calculate the correlator function in terms of hadronic and
quark--gluon degrees of freedom.

The contribution of the decuplet baryons to the correlation function
(\ref{eodt01}) is obtained in the following way,
\bea
\label{eodt05}
\Pi_{\mu\nu} = {1\over m_{\cal D}^2 - p^{\prime 2}} \lla 0 \vel \eta_\mu(0) \ver
{\cal D}(p^{\prime}) \rra \lla {\cal D}(p^{\prime}) \vel j_\nu^{el}(x) \ver {\cal
O}(p) \rra  + \cdots~,
\eea
where $\cdots$ refers to the contributions of the higher states with the same
quantum numbers as decuplet baryons.
The matrix element $\lla 0 \vel \eta_\mu(0) \ver {\cal D}(p^{\prime}) \rra$ is
determined as
\bea
\label{eodt06}
\lla 0 \vel \eta_\mu(0) \ver {\cal D}(p^{\prime}) \rra = \lambda_{\cal D}
u_\mu(p^{\prime})~,
\eea
where $\lambda_{\cal D}$ is the residue for the corresponding decuplet
baryon. The matrix element $\lla {\cal D}(p^{\prime}) \vel j_\nu^{el}(x) \ver
{\cal O}(p) \rra$ can be expressed in terms of three form factors $G_1$,
$G_2$ and $G_3$ with the help of Eq. (\ref{eodt02}). Summation over spins of
spin--3/2 baryons is defined as,
\bea
\label{eodt07}
\sum_s u_\mu(p^{\prime}\!,s) \bar{u}_\alpha(p^{\prime}\!,s) \es
-({p}^{\prime}\!\!\!\!/ + m_{\cal D}) \Bigg[ g_{\mu\alpha} - {1\over 3} \gamma_\mu
\gamma_\alpha - {2 p_\mu^{\prime} p_\alpha^{\prime} \over 3 m_{\cal D}^2} +
{p_\mu^{\prime} \gamma_\alpha - p_\alpha^{\prime} \gamma_\mu \over 3 m_{\cal D}}
\Bigg]~. 
\eea
The contributions of the decuplet baryons to the correlation function can be
obtained by substituting Eqs. (\ref{eodt02}) and (\ref{eodt03}) in Eq.
(\ref{eodt05}), from which we we get
\bea
\label{eodt08}
\Pi_{\mu\nu}(p,q) \es
- {1 \over m_{\cal D}^2 - p^{\prime 2}} \lambda_{\cal D}
({p}^{\prime}\!\!\!\!/ + m_{\cal D}) \Bigg[ g_{\mu\alpha} - {1\over 3} \gamma_\mu
- \gamma_\alpha - {2 p_\mu^{\prime} p_\alpha^{\prime} \over 3 m_{\cal D}^2} +
{p_\mu^{\prime} \gamma_\alpha - p_\alpha^{\prime} \gamma_\mu \over 3 m_{\cal D}}
\Bigg] \nnb \\
\cp \Bigg\{ G_1(Q^2) \Big(\bar{q}_\alpha \gamma_\nu +
\rlap/{q}g_{\alpha\nu}\Big) + {G_2(Q^2)\over 2} \Big[ -q_\alpha {\cal P}_\nu
+ (q\mcdot {\cal P}) g_{\alpha\nu} \Big] \nnb \\
\ar  G_3(Q^2) \Big[ q_\alpha q_\nu - q^2
g_{\alpha\nu} \Big] \Bigg\} \gamma_5 u(p)~.
\eea
It should be remarked here that Eq. (\ref{eodt04}) contains contributions
not only from the decuplet baryons, but also from  spin--parity $1^-/2$
baryons. Indeed, the matrix elements of the $\eta_\mu$ current sandwiched
between vacuum and the spin--parity $1^- /2$ state (denoted by ${\cal
D}^\ast$) is determined as
\bea
\label{eodt09}
\lla 0 \vel \eta_\mu (0) \ver {\cal D}^\ast (p^\prime) \rra = A (\gamma_\mu
m_{\ast} - 4 p_\mu^\prime ) u^\ast (p^\prime)~.
\eea
It follows from Eq. (\ref{eodt05}) that the unwanted spin--1/2 contribution
contains terms multiplied by $p^\prime$ or $\gamma_\mu$ at left, which all
must be eliminated. For this purpose an ordering procedure of Dirac matrices is
needed, as the result of which we also obtain the independent structures.
In this work we choose the ordering of the Dirac matrices as $\gamma_\mu
{p}^{\prime}\!\!\!\!/ \rlap/{q} \gamma_\nu \gamma_5$. After eliminating the
contributions of the spin--1/2 baryons, the correlation function takes the
following form:
\bea
\label{eodt10}
\Pi_{\mu\nu}(p,q) \es
- {1 \over m_{\cal D}^2 - p^{\prime 2}} \lambda_{\cal D}
({p}^{\prime}\!\!\!\!/ + m_{\cal D}) \Bigg\{ (- q_\mu \gamma_\nu +
\rlap/{q} g_{\mu\nu})  G_1(Q^2) \nnb \\
\ar \Big[ - q_\mu {\cal P}_\nu + (q\mcdot
{\cal P}) g_{\mu\nu} \Big] {G_2(Q^2) \over 2}
+ \Big[q_\mu q_\nu - q^2 g_{\mu\nu} \Big] G_3(Q^2) \Bigg\} \gamma_5 u(p)~.
\eea
The above expression for the correlation function can be decomposed into
contributions of various Lorentz structures, any three of which can be used
in the numerical calculations of the form factors $G_1$, $G_2$ and $G_3$.
In the present work we choose the structures ${p}^{\prime}\!\!\!\!/
\rlap/{q} \gamma_5 g_{\mu\nu}$, ${p}^{\prime}\!\!\!\!/ 
\gamma_5 q_\mu p_\nu^\prime$ and
${p}^{\prime}\!\!\!\!/ \gamma_5 q_\mu q_\nu$, in determination of the form
factors $G_1$, $G_2$ and $\ds{G_2\over 2} - G_3$, respectively. The
invariant functions in the correlation function corresponding to the
structures ${p}^{\prime}\!\!\!\!/         
\rlap/{q} \gamma_5 g_{\mu\nu}$, $\rlap/{q} \gamma_5 q_\mu p_\nu^\prime$ and      
${p}^{\prime}\!\!\!\!/ \gamma_5 q_\mu q_\nu$ are given as,
\bea
\label{eodt11}
\Pi^{(1)} \es - {\lambda_{\cal D} \over m_{\cal D}^2 - p^{\prime 2}}
G_1(Q^2)~, \nnb \\
\Pi^{(2)} \es {\lambda_{\cal D} \over m_{\cal D}^2 - p^{\prime 2}} 
G_2(Q^2)~, \nnb \\
\Pi^{(3)} \es {\lambda_{\cal D} \over m_{\cal D}^2 - p^{\prime 2}} 
\Bigg[{G_2(Q^2)\over 2} - G_3(Q^2)\Bigg]~,
\eea
respectively. In order to construct sum rules for the form factors $G_1$,
$G_2$ and $\ds{G_2\over 2} - G_3$ we need the corresponding expressions of
the correlation function from the QCD side, in which distribution  
amplitudes (DAs) of the octet baryons are contained. Since the $\gamma^* N \to
\Delta$ transition has already been investigated in the framework of the QCD sum
rules in \cite{Rodt13}, we restrict our analysis to the consideration of the
$\gamma^* \Sigma \to \Sigma^\ast$, $\gamma^* \Lambda \to \Sigma^\ast$,
and $\gamma^* \Xi \to \Xi^\ast$ transitions.

The distribution amplitudes of the octet baryons are defined as a matrix
element of the three--quark operator between one of the members of the octet
baryon and vacuum, i.e.,
\bea
\label{nolabel}
\varepsilon^{abc} \lla 0 \vel q_{1\alpha}^a(a_1x)
q_{2\beta}^b(a_2x)q_{3\gamma}^c(a_3x) \ver {\cal O}(p) \rra~, \nnb
\eea
where $a$, $b$, $c$ are the color indices; $\alpha$, $\beta$ and $\gamma$ are the
Lorentz indices; and $a_i$ are positive numbers satisfying $a_1+a_2+a_3=1$.

Lorentz covariance, together with spin and parity of the baryons, imposes that
the general Lorentz composition of the matrix element is
\bea
\label{eodt12}
4 \varepsilon^{abc} \lla 0 \vel q_{1\alpha}^a(a_1x)
q_{2\beta}^b(a_2x)q_{3\gamma}^c(a_3x) \ver {\cal O}(p) \rra = \sum_i {\cal
F}_i \Gamma_{\alpha\beta}^{1i} \Big(\Gamma^{2i} {\cal O}(p)\Big)_\gamma~,
\eea
where $\Gamma^{1(2)i}$ are certain Dirac matrices, and ${\cal F}_i = {\cal
S}_i$, ${\cal P}_i$, ${\cal A}_i$, ${\cal V}_i$ and ${\cal T}_i$ are the
DAs having no definite twists. This decomposition in terms of the functions
${\cal F}_i$ can be found in \cite{Rodt15}, and for completeness it is
presented in Appendix A.

The DAs with definite twist are given as,
\bea
\label{eodt13}
4 \varepsilon^{abc} \la 0 \vel q_{1\alpha}^a(a_1x)  q_{2\beta}^b(a_2x)
q_{3\gamma}^c(a_3x) \ver {\cal O}(p) \ra = \sum_i F_i
\Gamma^{\prime 1i}_{\alpha\beta} \Big(\Gamma^{\prime 2i} {\cal O}(p)\Big)_\gamma~,
\eea
where $F_i=S_i,P_i,A_i,V_i,A_i$ and $T_i$. The relations among these two sets of
DAs are given as,
\bea
\label{nolabel}
\begin{array}{ll}
{\cal S}_1 = S_1~,& (2 P \mcdot x) \, {\cal S}_2 = S_1 - S_2~, \nnb \\
{\cal P}_1 = P_1~,& (2 P \mcdot x) \, {\cal P}_2 = P_2 - P_1~, \nnb \\
{\cal V}_1 = V_1~,& (2 P \mcdot x) \, {\cal V}_2 = V_1 - V_2 - V_3~, \nnb \\
2{\cal V}_3 = V_3~,& (4 P \mcdot x) \, {\cal V}_4 =
- 2 V_1 + V_3 + V_4 + 2 V_5~,\nnb \\
(4 P\mcdot x) \, {\cal V}_5 = V_4 - V_3~,&
(2 P \mcdot x)^2 \, {\cal V}_6 = - V_1 + V_2 + V_3 + V_4 + V_5 - V_6~, \nnb \\
{\cal A}_1 = A_1~,& (2 P \mcdot x) \, {\cal A}_2 = - A_1 + A_2 - A_3~, \nnb \\
2 {\cal A}_3 = A_3~,&
(4 P \mcdot x) \, {\cal A}_4 = - 2 A_1 - A_3 - A_4 + 2 A_5~, \nnb \\
(4 P \mcdot x) \, {\cal A}_5 = A_3 - A_4~,&
(2 P \mcdot x)^2 \, {\cal A}_6 = A_1 - A_2 + A_3 + A_4 - A_5 + A_6~, \nnb \\
{\cal T}_1 = T_1~, & (2 P \mcdot x) \, {\cal T}_2 = T_1 + T_2 - 2T_3~, \nnb \\
2 {\cal T}_3 = T_7~,& (2 P \mcdot x) \, {\cal T}_4 = T_1 - T_2 - 2 T_7~, \nnb \\
(2 P\mcdot x) \, {\cal T}_5 = - T_1 + T_5 + 2 T_8~,&
(2 P \mcdot x)^2 \, {\cal T}_6 = 2 T_2 - 2 T_3 - 2 T_4 + 2 T_5 + 2 T_7 + 2 T_8~, \nnb \\
(4P \mcdot x) \, {\cal T}_7 = T_7 - T_8~, &
(2 P\mcdot x)^2 \, {\cal T}_8 = - T_1 + T_2 + T_5 - T_6 + 2 T_7 + 2 T_8~. \nnb
\end{array}
\eea
Explicit expressions of DAs ${\cal S}_i$, ${\cal P}_i$, ${\cal A}_i$, ${\cal V}_i$
and ${\cal T}_i$ at the leading order of conformal spin expansion
can be found in \cite{Rodt08,Rodt09,Rodt10,Rodt11}.

After lengthy calculations, we obtain the invariant functions $\Pi_i$ 
for the $\gamma^\ast~octet \to
decuplet$ transitions. Schematically, the expressions for $\Pi_i$ can be
written in terms of the functions $\rho_2(x)$, $\rho_4(x)$, and $\rho_6(x)$
as follows:
\bea
\label{eodt14}
\Pi = N \int_0^1 dx \Bigg\{ {\rho_2(x) \over (q-px)^2} +
{\rho_4(x) \over (q-px)^4} +
{\rho_6(x) \over (q-px)^6} \Bigg\}~.
\eea
Explicit expressions of the functions $\rho_2(x)$, $\rho_4(x)$, and  $\rho_6(x)$
for the considered transitions are given in Appendix B.

Equating the invariant functions $\Pi_i(Q^2),~(i=1,2,3)$ from the QCD and
hadronic sides, and performing Borel transformation over the variable
$-(p-q)^2$, we obtain the following sum rules for the form factors
$G_i(Q^2)$:

\bea
\label{eodt15}
G_i^\alpha(Q^2) \es {N_\alpha \over 2 \lambda_{\cal D}} e^{m_{\cal D}^2/M^2}
\Bigg\{ \int_{x_0}^1 dx \Bigg(
- {(\rho_2(x))_i^\alpha\over x} + {(\rho_4(x))_i^\alpha\over x^2 M^2} -
- {(\rho_6(x))_i^\alpha\over 2 x^3 M^4} \Bigg)
e^{-{\bar{x} Q^2 \over x M^2} - {\bar{x} m_{\cal O}^2 \over M^2} } \nnb \\
\ar \Bigg[ {(\rho_4 (x_0))_i^\alpha \over Q^2 + x_0^2 m_{\cal O}^2} - {1\over 2 x_0}
{(\rho_6(x_0))_i^\alpha \over (Q^2 + x_0^2 m_{\cal O}^2) M^2} \nnb \\
\ar {1\over 2} {x_0^2 \over (Q^2 + x_0^2 m_{\cal O}^2)} \Bigg(
{d\over dx_0} {(\rho_6(x_0))_i^\alpha \over x_0 (Q^2 + x_0^2 m_{\cal O}^2) M^2}
\Bigg) \Bigg]e^{-s_0/ M^2}
\Bigg\}~,
\eea
where $i=1,2,3$ correspond to the form factors $G_1(Q^2)$,
$G_2(Q^2)$, $\ds{G_2(Q^2)\over 2}-G_3(Q^2)$, and $\alpha$ corresponds to any
member of the decuplet.
Here $M^2$ is the square of the Borel mass parameter, $s={\ds {\bar{x}\over
x}} Q^2 + \bar{x} m_{\cal O}^2$,  $x_0$ is the solution of the equation
$s=s_0$, $m_{\cal O}$ and $m_{\cal D}$ are masses of the members of the
octet and decuplet baryons, respectively, and $\bar{x}=1-x$.

As has already been noted, the $\gamma^\ast N \to \Delta$ transition is
studied within the light cone QCD sum rules in \cite{Rodt13}; hence, we do
not consider it in this work.

The sum rules needed in determining the three form factors $G_1(Q^2)$,
$G_2(Q^2)$, and $\ds{G_2(Q^2)\over 2}-G_3(Q^2)$
are given in Eq. (\ref{eodt15}). The form factors
$G_1(Q^2)$ and $G_2(Q^2)$ are obtained from Eqs. (\ref{eodt15}).
With the help of these three form factors, we finally
rewrite our results in terms of the magnetic dipole $G_M(Q^2)$, electric
quadrupole $G_E(Q^2)$, and Coulomb quadrupole $G_C(Q^2)$ form factors.

\section{Numerical Analysis}

In this section we present our numerical result for the magnetic dipole
$G_M(Q^2)$, electric quadrupole $G_E(Q^2)$, and Coulomb quadrupole
$G_C(Q^2)$ form factors. It follows from the explicit expressions of the
sum rules for these form factors that in order to determine these form
factors, DAs of the octet baryons are needed. A few words about the DAs of
the octet baryons are in order. The distribution amplitudes of the nucleon
within the next--to--leading order in conformal spin are calculated in
\cite{Rodt15}. These results are then extended to next--to--next
order by calculating DAs with twist--3 in conformal spin in \cite{Rodt16}.
In the present work, we shall use DAs of the $\Lambda$, $\Sigma$, and $\Xi$
octet baryons which are given in \cite{Rodt17,Rodt18,Rodt19,Rodt20},
and exclude these contributions, which have not yet been
calculated. 

The nonperturbative parameters $f_{\cal O}$, $\lambda_1$, $\lambda_2$
and $\lambda_3$ appearing in the expressions of the DAs are given in
\cite{Rodt17,Rodt18,Rodt19,Rodt20}, which are determined from the analysis of the
two--point correlation function.

\bea
\label{nolabel}
f_\Xi \es (9.9 \pm 0.4)\times 10^{-3}~GeV^2~, \nnb \\
\lambda_1 \es -(2.1 \pm 0.1)\times 10^{-2}~GeV^2~, \nnb \\ 
\lambda_2 \es (5.2 \pm 0.2)\times 10^{-2}~GeV^2~, \nnb \\
\lambda_3 \es (1.7 \pm 0.1)\times 10^{-2}~GeV^2~, \nnb \\ \nnb \\
f_\Sigma \es (9.4 \pm 0.4)\times 10^{-3}~GeV^2~, \nnb \\ 
\lambda_1 \es -(2.5 \pm 0.1)\times 10^{-2}~GeV^2~, \nnb \\
\lambda_2 \es (4.4 \pm 0.1)\times 10^{-2}~GeV^2~, \nnb \\  
\lambda_3 \es (2.0 \pm 0.1)\times 10^{-2}~GeV^2~, \nnb \\ \nnb \\
f_\Lambda \es (6.0 \pm 0.3)\times 10^{-3}~GeV^2~, \nnb \\ 
\lambda_1 \es (1.0 \pm 0.3)\times 10^{-2}~GeV^2~, \nnb \\
\vel \lambda_2 \ver \es (0.83 \pm 0.05)\times 10^{-2}~GeV^2~, \nnb \\  
\vel \lambda_3 \ver \es (0.83 \pm 0.05)\times 10^{-2}~GeV^2~. \nnb
\eea

Furthermore, in calculating the form factors from the QCD sum rules analysis
we need to find the working regions of the Borel mass parameter $M^2$ and
continuum threshold $s_0$. The continuum threshold is not completely
arbitrary, and depends on the energy of the first excited states with the
same quantum numbers. In our numerical calculations we will use
$s_0=4.0~GeV^2$ which is obtained from the mass sum rule analysis
\cite{Rodt21}. The working region of $M^2$ can be found by using the
following criteria. 
\begin{itemize}
\item The lower limit of $M^2$ is determined by demanding
that the contributions coming from the higher states and continuum are less
than half of the total result. 
\item The upper bound can be obtained by imposing the conditions required by
the operator product expansion.
\end{itemize}
It follows from the numerical calculations that these two conditions are
both fulfilled in the region $1.5 \le M^2 \le 3.5~GeV^2$.

In the expressions of the form factors we see that the
residues of the decuplet baryons are needed. These residues are calculated
in \cite{Rodt09}, which we shall use in further numerical analysis.

We have already noted, from an experimental point of view, that the multipole form
factors constitute a more suitable set compared to the form factors
$G_1$, $G_2$ and $G_1$. For this reason, we will present the results for
multipole form factors.

In Figs. (1)--(3), we present the $Q^2$ dependence of the magnetic
dipole $G_M(Q^2)$, electric quadrupole $G_E(Q^2)$, and Coulomb quadrupole
$G_C(Q^2)$ form factors for the $\gamma^\ast \Sigma^+ \to \Sigma^{\ast +}$
transition, at $s_0=4.0~GeV^2$ and at several different values of the Borel
mass parameter $M^2$. In the numerical
calculations $Q^2$ is varied in the region $1.0 \le Q^2 \le
8.0~GeV^2$, because in this region the higher twist contributions and
the higher states and continuum contributions are small.

It follows from these figures that the electric quadrupole form factor
$G_E(Q^2)$ is small compared to the magnetic dipole form factor $G_M(Q^2)$.

Figures (4)--(8) depict the dependence of the magnetic dipole
form factor $G_M(Q^2)$ on $Q^2$, at $s_0=4.0~GeV^2$ and
at several different values of $M^2$, for the  
$\gamma^\ast \Lambda \to \Sigma^{\ast 0}$, $\gamma^\ast \Sigma^0 \to
\Sigma^{\ast 0}$, $\gamma^\ast \Sigma^- \to \Sigma^{\ast -}$, $\gamma^\ast
\Xi^- \to \Xi^{\ast -}$, and $\gamma^\ast \Xi^0 \to \Xi^{\ast 0}$,
transitions,
respectively. We see from these figures that the magnitude of the magnetic
dipole form factors $G_M(Q^2)$ for the $\gamma^\ast \Xi^0 \to \Xi^{\ast 0}$
and $\gamma^\ast \Sigma^+ \to \Sigma^{\ast +}$ transitions are practically
equal. We also observe that the value of $G_M(Q^2)$ is quite small for the
$\gamma^\ast \Sigma^- \to \Sigma^{\ast -}$ transition  and is small for the
$\gamma^\ast \Xi^- \to \Xi^{\ast -}$ transition. So, had the magnitudes of
the magnetic dipole form factor $G_M(Q^2)$ been classified, they could be
ordered as large for $\gamma^\ast \Sigma^+ \to \Sigma^{\ast +}$,
moderate for $\gamma^\ast \Lambda \to \Sigma^{\ast 0}$, and small for the
$\gamma^\ast \Sigma^- \to \Sigma^{\ast -}$ and $\gamma^\ast  
\Xi^- \to \Xi^{\ast -}$ transitions. These results can be explained as a
consequence of $U$--spin symmetry \cite{Rodt22}.

In order to get an idea about the order of the violation of $U$--spin
symmetry, we consider the ratio
$\vel ( G_M^{\Sigma^+} / G_M^{\Xi^0}) - 1 \ver$. In the case of $U$--spin
symmetry this quantity should be equal to zero. Our numerical results show
that this quantity is about 0.3, i.e., the violation of $U$--spin
symmetry is about 30\%. On the other hand, in the case of the $\gamma^\ast \Sigma^- \to
\Sigma^{\ast -}$ and $\gamma^\ast \Xi^- \to \Xi^{\ast -}$ transitions, the
above--considered ratio is approximately equal to 2.0, which is an
identification of the highly broken $U$--spin symmetry. It should be
remembered here that, the values of $G_M(Q^2)$ for these transitions are quite
small and they are very sensitive to the values of the input parameters.

For completeness, we can also compare our results with the quark model predictions
by means of the ratios $R_1 = (G_M^{\Sigma^+})_{quark} /
(G_M^{\Sigma^+})_{our}$ and $R_2 = (G_M^{\Xi^0})_{quark} /
(G_M^{\Xi^0})_{our}$. Our analysis shows that these ratios vary in the
range 1.3 to 1.4 in accordance with their dependence on $Q^2$. In other
words, the difference among our predictions and those of the quark model results   
is about 40\%.

The $U$--spin symmetry allows us to obtain the relations among the
form factors of $\gamma^\ast \Lambda \to \Sigma^{\ast 0}$ and
$\gamma^\ast N \to \Delta$ transitions, the latter of which has already been
measured in the experiments. One can easily find that the form factors in
the aforementioned transitions differ from each other by a factor of
$\sqrt{3/4}$. Using this result we can compare our predictions on the
multipole form factors with the predictions of \cite{Rodt13}, in which the
magnetic dipole $G_M(Q^2)$ for the $\gamma^\ast N \to \Delta$ transition is
calculated within the framework of the light cone QCD sum rules. From a
comparison of our result on $G_M(Q^2)$ with the result given in
\cite{Rodt13}, we see that the prediction of both works are very close to each
other. 

The results for the electric quadrupole $G_E(Q^2)$ and Coulomb quadrupole
$G_C(Q^2)$ form factors can be summarized as follows. From our numerical
results we observe that, for
the $\gamma^\ast \Xi^0 \to \Xi^{\ast 0}$ transition only, the electric
quadrupole $G_E(Q^2)$ form factor changes its sign around $Q^2\simeq
2.5~GeV^2$. In the transitions $\gamma^\ast \Sigma^+ \to \Sigma^{\ast +}$,
$\gamma^\ast \Sigma^0 \to \Sigma^{\ast 0}$, $\gamma^\ast \Lambda \to
\Sigma^{\ast 0}$, $\gamma^\ast
\Xi^- \to \Xi^{\ast -}$ and $\gamma^\ast \Sigma^- \to \Sigma^{\ast -}$,
the values of the electric quadrupole $G_E(Q^2)$ form
factors are negative in the range $0.0 \le Q^2 \le 8.0~GeV^2$.
Note that the maximum value of the electric quadrupole
$G_E(Q^2)$ is about $0.06$.
 
The behavior of the Coulomb quadrupole $G_C(Q^2)$ form factor for the
$\gamma^\ast \Sigma^+ \to \Sigma^{\ast +}$ and $\gamma^\ast \Xi^0 \to
\Xi^{\ast 0}$ transitions is quite similar and the magnitude of $G_C(Q^2)$ 
and their values are close to each other. The maximum values of the Coulomb
quadrupole $G_C(Q^2)$ form factor for the $\gamma^\ast \Sigma^0 \to
\Sigma^{\ast 0}$ and $\gamma^\ast \Lambda \to \Sigma^{\ast 0}$ transitions
are very close to each other. they are about $0.10$--$0.15$.  For the
$\gamma^\ast \Sigma^- \to \Sigma^{\ast -}$ and $\gamma^\ast
\Xi^- \to \Xi^{\ast -}$ transitions, the values of $G_C(Q^2)$ are small.
Their maximum value is about $0.012$.

A comparison of our results on magnetic dipole form factor  $G_M(Q^2)$
with those given in \cite{Rodt11} would be quite useful.
Our predictions on the magnitude of $G_M(Q^2)$ for the $\gamma^\ast \Xi^0 \to \Xi^{\ast
0}$, $\gamma^\ast \Lambda \to \Sigma^{\ast 0}$, $\gamma^\ast \Sigma^0 \to
\Sigma^{\ast 0}$, and $\gamma^\ast \Sigma^+ \to \Sigma^{\ast +}$ transitions
are smaller compared to the ones obtained from the quark model, while it is
contrary to this case for the $\gamma^\ast \Xi^- \to \Xi^{\ast -}$ transition.

We finally study the $Q^2$ dependence of the ratios $R_{EM}$ and $R_{SM}$
[see Eq. (\ref{eodt04})].

In Figs (9) and (10) we study the $Q^2$ dependence of the ratios
$R_{EM}$ and $R_{SM}$ on $Q^2$, respectively. We see from
Fig. (9) that the predictions for $R_{EM}$ for the $\gamma^\ast
\Sigma^- \to \Sigma^{\ast -}$ and $\gamma^\ast \Xi^- \to \Xi^{\ast -}$
transitions are close to each other and the values of $R_{EM}$ for these
transitions range in the regions $0.4$--$0.5$ and $0.50$--$0.65$,
respectively. For all other transitions the maximum value $R_{EM}$ is about
$0.2$. It should also be noted that the values of $R_{EM}$ for the
$\gamma^\ast \Sigma^+ \to \Sigma^{\ast +}$ and $\gamma^\ast \Sigma^0 \to
\Sigma^{\ast 0}$ transitions do not change considerably, while
for the $\gamma^\ast \Xi^0 \to \Xi^{\ast 0}$ and $\gamma^\ast \Lambda \to
\Sigma^{\ast 0}$ transitions $R_{EM}$ varies from $0.0$ at $Q^2=1.0~GeV^2$ to
$0.2$ at $Q^2=8.0~GeV^2$.

As far as $R_{SM}$ is considered, our results can be summarized as follows.

\begin{itemize}

\item The result of $R_{SM}$ for the case of the $\gamma^\ast \Sigma^- \to
\Sigma^{\ast -}$ transition shows that $R_{SM}$ is practically very small
when $Q^2$ varies in the domain $1.0 \le Q^2 \le 8.0~GeV^2$.

\item  Similar to the previous case, the value of $R_{SM}$ is also very
small for the $\gamma^\ast \Xi^- \to \Xi^{\ast -}$ transition, which
can be attributed to the $U$--spin symmetry. The small difference in the results
has its roots in the violation of $SU(3)$ symmetry.

\item The values of $R_{SM}$ for the $\gamma^\ast \Lambda \to \Sigma^{\ast
0}$ transition run in the range $0.2$--$0.3$.

\item Furthermore, the behavior of $R_{SM}$ for the $\gamma^\ast \Sigma^+
\to \Sigma^{\ast +}$ and $\gamma^\ast \Xi^0 \to \Xi^{\ast 0}$ transitions
seems to be practically independent of $Q^2$, and is about $0.25$ and
$0.35$, respectively.

\item As we consider the $\gamma^\ast \Sigma^0 \to \Sigma^{\ast 0}$
transition, we observe that it exhibits sensitivity to the variation of
$Q^2$, and $R_{SM}$ changes in the range $0.35$--$0.50$.

\end{itemize}

Determination of the multipole form factors for the $\gamma^\ast~octet \to
decuplet$ transitions from future experiments would be quite valuable in
checking the predictions of various theoretical models and for choosing the
``right" model of hadrons.
In this sense our
predictions might shed light on a deeper understanding of the inner structure
of hadrons. 

The results presented in this work can be improved by extending the
calculations for the DAs to the next--to--leading order of the conformal
spin and by taking ${\cal O}(\alpha_s)$ corrections into account.

In conclusion, in this work we study the multipole form factors, namely, 
magnetic dipole $G_M(Q^2)$, electric quadrupole $G_E(Q^2)$, and Coulomb
quadrupole $G_C(Q^2)$ form factors, describing the octet to decuplet
electromagnetic transitions within the light cone QCD sum
rules. We compare our results on the magnetic dipole $G_M(Q^2)$ form factor
with the predictions of the spectator quark model \cite{Rodt11} and see
that our results are smaller in magnitude than those predicted in
\cite{Rodt11}, except for the $\gamma^\ast \Xi^- \to \Xi^{\ast -}$
transition. We next study the $Q^2$ dependence of the ratios of electric
quadrupole $G_E(Q^2)$ and Coulomb quadrupole $G_C(Q^2)$ form factors to the
magnetic dipole $G_M(Q^2)$ form factor.

\newpage


\section*{Appendix A}
\setcounter{equation}{0}
For completeness, in this appendix we present the general Lorentz decomposition of
the matrix element of the three--quark operators between the vacuum and the
octet baryon states in terms of the DAs \cite{Rodt15}.

\bea
\label{appendixA}
\lefteqn{ 4 \lla 0 \vel \varepsilon^{ijk} u_\alpha^i(a_1 x) u_\beta^j(a_2 x) d_\gamma^k(a_3 x)
\ver {\cal O}(p) \rra = } \nnb \\
&&  
{\cal S}_1 m_{\cal O} C_{\alpha \beta} \left(\gamma_5 {\cal O}\right)_\gamma + 
{\cal S}_2 m_{\cal O}^2 C_{\alpha \beta} \left(\!\not\!{x} \gamma_5 {\cal O}\right)_\gamma + 
{\cal P}_1 m_{\cal O} \left(\gamma_5 C\right)_{\alpha \beta} {\cal O}_\gamma + 
{\cal P}_2 m_{\cal O}^2 \left(\gamma_5 C \right)_{\alpha \beta} \left(\!\not\!{x} {\cal O}\right)_\gamma 
\nnb \\
\ar 
\left({\cal V}_1+\frac{x^2m_{\cal O}^2}{4}{\cal V}_1^M \right)
\left(\!\not\!{p}C \right)_{\alpha \beta} \left(\gamma_5 {\cal O}\right)_\gamma + 
{\cal V}_2 m_{\cal O} \left(\!\not\!{p} C \right)_{\alpha \beta} \left(\!\not\!{x} \gamma_5 {\cal O}\right)_\gamma  + 
{\cal V}_3 m_{\cal O}  \left(\gamma_\mu C \right)_{\alpha \beta}\left(\gamma^{\mu} \gamma_5 {\cal O}\right)_\gamma 
\nnb \\ 
\ar
{\cal V}_4 m_{\cal O}^2 \left(\!\not\!{x}C \right)_{\alpha \beta} \left(\gamma_5 {\cal O}\right)_\gamma +
{\cal V}_5 m_{\cal O}^2 \left(\gamma_\mu C \right)_{\alpha \beta} \left(i \sigma^{\mu\nu} x_\nu \gamma_5 
{\cal O}\right)_\gamma 
+ {\cal V}_6 m_{\cal O}^3 \left(\!\not\!{x} C \right)_{\alpha \beta} \left(\!\not\!{x} \gamma_5 {\cal O}\right)_\gamma  
\nnb \\ 
\ar 
\left({\cal A}_1+\frac{x^2m_{\cal O}^2}{4}{\cal A}_1^M\right)
\left(\!\not\!{p}\gamma_5 C \right)_{\alpha \beta} {\cal O}_\gamma + 
{\cal A}_2 m_{\cal O} \left(\!\not\!{p}\gamma_5 C \right)_{\alpha \beta} \left(\!\not\!{x} {\cal O}\right)_\gamma  + 
{\cal A}_3 m_{\cal O}  \left(\gamma_\mu \gamma_5 C \right)_{\alpha \beta}\left( \gamma^{\mu} {\cal O}\right)_\gamma 
\nnb \\ 
\ar
{\cal A}_4 m_{\cal O}^2 \left(\!\not\!{x} \gamma_5 C \right)_{\alpha \beta} {\cal O}_\gamma +
{\cal A}_5 m_{\cal O}^2 \left(\gamma_\mu \gamma_5 C \right)_{\alpha \beta} \left(i \sigma^{\mu\nu} x_\nu  
{\cal O}\right)_\gamma 
+ {\cal A}_6 m_{\cal O}^3 \left(\!\not\!{x} \gamma_5 C \right)_{\alpha \beta} \left(\!\not\!{x} {\cal O}\right)_\gamma  
\nnb \\
\ar
\left({\cal T}_1+\frac{x^2m_{\cal O}^2}{4}{\cal T}_1^M\right)
\left(p^\nu i \sigma_{\mu\nu} C\right)_{\alpha \beta} \left(\gamma^\mu\gamma_5 {\cal O}\right)_\gamma 
+ 
{\cal T}_2 m_{\cal O} \left(x^\mu p^\nu i \sigma_{\mu\nu} C\right)_{\alpha \beta} \left(\gamma_5 {\cal O}\right)_\gamma 
\nnb \\
\ar 
 {\cal T}_3 m_{\cal O} \left(\sigma_{\mu\nu} C\right)_{\alpha \beta} \left(\sigma^{\mu\nu}\gamma_5 {\cal O}\right)_\gamma 
+ {\cal T}_4 m_{\cal O} \left(p^\nu \sigma_{\mu\nu} C\right)_{\alpha\beta} \left(\sigma^{\mu\rho} x_\rho \gamma_5 {\cal O}\right)_\gamma 
\nnb \\
\ar {\cal T}_5 m_{\cal O}^2 \left(x^\nu i \sigma_{\mu\nu} C\right)_{\alpha \beta} \left(\gamma^\mu\gamma_5 {\cal O}\right)_\gamma 
+ 
{\cal T}_6 m_{\cal O}^2 \left(x^\mu p^\nu i \sigma_{\mu\nu} C\right)_{\alpha \beta} \left(\!\not\!{x} \gamma_5 {\cal O}\right)_\gamma  
\nnb \\
\ar {\cal T}_{7} m_{\cal O}^2 \left(\sigma_{\mu\nu} C\right)_{\alpha \beta} \left(\sigma^{\mu\nu} \!\not\!{x} \gamma_5 {\cal O}\right)_\gamma
+ {\cal T}_{8} m_{\cal O}^3 \left(x^\nu \sigma_{\mu\nu} C\right)_{\alpha \beta} \left(\sigma^{\mu\rho} x_\rho \gamma_5 {\cal O}\right)_\gamma ~,
\nnb
\eea
where $C$ is the charge conjugation operator, and ${\cal O}$ represents the
octet baryon with momentum $p$. 

\newpage

\section*{Appendix B}  
\setcounter{equation}{0}

In this appendix we present the expressions for the functions
$\rho_2$, $\rho_4$ and $\rho_6$
which appear in the sum rules for $G_1(Q^2)$,
$G_2(Q^2)$, and $\ds{G_2(Q^2)\over 2} - G_3(Q^2)$, for the $\gamma^\ast
\Sigma^+ \to \Sigma^{\ast +}$ and $\gamma^\ast
\Sigma^0 \to \Sigma^{\ast 0}$ transitions.

\section*{$\gamma^\ast \Sigma^+ \to \Sigma^{\ast +}$ transition}
\bea
\label{Sigma+}
%
\left( \rho_4\right)_1^{\Sigma^{\ast +}} (x)\es
4 e_{q_3} m_{\cal O}^2 \; \widehat{\!\widehat{B}}_6 (x)
- 8 e_{q_2} m_{\cal O} m_{q_2} \widetilde{B}_2 (x)
- 8 e_{q_3} m_{\cal O} m_{q_3} \widehat{B}_4 (x) \nnb \\
\ek 8 e_{q_2} m_{\cal O}^2 \int_0^{\bar{x}} \,dx_1 \Big[A_1^M - T_1^M\Big](x_1,x,1-x_1-x)
 \nnb \\
\ek 8 e_{q_3} m_{\cal O}^2 \int_0^{\bar{x}}dx_1\, T_1^M (x_1,1-x_1-x,x)
\nnb\\ \nnb \\
%
\left( \rho_2\right)_1^{\Sigma^{\ast +}} (x)\es
-8 e_{q_2} \int_0^{\bar{x}} \,dx_1 \Big[A_1-T_1 \Big] (x_1,x,1-x_1-x)
\nnb \\
\ek 8 e_{q_3} \int_0^{\bar{x}}dx_1\,  T_1 (x_1,1-x_1-x,x)
\nnb\\ \nnb \\
%
\left( \rho_6\right)_2^{\Sigma^{\ast +}} (x)\es
-64 e_{q_1} m_{\cal O}^3 (1-x) x^2 \;\check{\!\check{C}}_6 (x) \nnb \\
\ar 16 e_{q_2} m_{\cal O}^2 x \Big[4 m_{\cal O} (1-x) x \, (2 \; \widetilde{\!\widetilde{B}}_8 -
\;\widetilde{\!\widetilde{C}}_6 ) - m_{q_2} \;\widetilde{\!\widetilde{B}}_6
\Big] (x)
\nnb \\
\ar 16 e_{q_3} m_{\cal O}^2 x \Big[ 2 m_{\cal O} (1-x) x \,(2 \; \widehat{\!\widehat{B}}_8
- 2 \; \widehat{\!\widehat{C}}_6 + \widehat{\!\widehat{D}}_6 ) +
m_{q_3} \widehat{\!\widehat{B}}_6 \Big] (x) \nnb \\ \nnb \\
%
\left( \rho_4\right)_2^{\Sigma^{\ast +}} (x)\es
-8 e_{q_1} m_{\cal O} (1-2 x) x \check{C}_2 \nnb \\
\ek 8 e_{q_2} m_{\cal O} x \Big[ 2 \widetilde{B}_2 - (1-2 x)  (2 \widetilde{B}_4 -
\widetilde{C}_2) + \widetilde{D}_2 \Big] (x) \nnb \\
\ar 8 e_{q_3} m_{\cal O} x \Big[ 2 (1-x) \widehat{B}_4 + x ( 2 \widehat{C}_2 -    
\widehat{D}_2) \Big] (x) \nnb \\ \nnb \\
%
\left( \rho_6\right)_3^{\Sigma^{\ast +}} (x)\es
64 e_{q_1} m_{\cal O}^3 (1-x)^2 x \; \check{\!\check{C}}_6 (x) \nnb \\
\ek 64 e_{q_2} m_{\cal O}^3 (1-x)^2 x \Big[ 2 \; \widetilde{\!\widetilde{B}}_8 -
\widetilde{\!\widetilde{C}}_6 \Big] (x) \nnb \\
\ek 32 e_{q_3} m_{\cal O}^3 (1-x)^2 x \Big[ 2 \; \widehat{\!\widehat{B}}_8 -
2 \; \widehat{\!\widehat{C}}_6  + \widehat{\!\widehat{D}}_6 \Big] (x)
\nnb \\ \nnb \\
%
\left( \rho_4\right)_3^{\Sigma^{\ast +}} (x)\es
- 16 e_{q_1} m_{\cal O} (1-x) x \check{C}_2 (x) \nnb \\
\ar 16 e_{q_2} m_{\cal O} (1-x) \Big[ \widetilde{B}_2 + x ( 2 \widetilde{B}_4 -        
\widetilde{C}_2 ) \Big] (x) \nnb \\
\ar 8 e_{q_3} m_{\cal O} (1-x) \Big[ \widehat{B}_2 - (1-2 x) \widehat{B}_4 +
x (2 \widehat{C}_2 - \widehat{D}_2 ) \Big] (x)~, \nnb
\eea
where $q_1=q_2=u$, $q_3=s$.

The expressions for
the functions $\rho_2$, $\rho_4$ and $\rho_6$ describing the $\gamma^\ast
\Xi^0 \to \Xi^{\ast 0}$ and $\gamma^\ast \Xi^- \to \Xi^{\ast -}$ transitions
can be obtained from the corresponding results of the $\gamma^\ast \Sigma^+
\to \Sigma^{\ast +}$ transition by making the replacements $ u \lrar s$ (for
the $\gamma^\ast \Xi^0 \to \Xi^{\ast 0}$), and $ u \to s$, $ s \to d$ (for 
the $\gamma^\ast \Xi^- \to \Xi^{\ast -}$).

\section*{$\gamma^\ast \Sigma^0 \to \Sigma^{\ast 0}$ transition}

\bea
%
\left( \rho_4\right)_1^{\Sigma^{\ast 0}} (x)\es
4 e_{q_3} m_{\cal O}^2 \; \widehat{\!\widehat{B}}_6 (x)
- 4 e_{q_1} m_{q_1} m_{\cal O} \check{B}_2 (x)
- 4 e_{q_2} m_{\cal O} m_{q_2} \widetilde{B}_2 (x)
- 8 e_{q_3} m_{\cal O} m_{q_3} \widehat{B}_4 (x) \nnb \\
\ar 4 e_{q_1} m_{\cal O}^2 \int_0^{\bar{x}} \,dx_3 \Big[A_1^M +                        
T_1^M\Big](x,1-x-x_3,x_3) \nnb \\
\ek 4 e_{q_2} m_{\cal O}^2 \int_0^{\bar{x}} \,dx_1 \Big[A_1^M - T_1^M\Big](x_1,x,1-x_1-x)
 \nnb \\
\ek 8 e_{q_3} m_{\cal O}^2 \int_0^{\bar{x}}dx_1\, T_1^M (x_1,1-x_1-x,x)
\nnb\\ \nnb \\
%
\left( \rho_2\right)_1^{\Sigma^{\ast 0}} (x)\es
    4 e_{q_1} \int_0^{\bar{x}} \,dx_3 \Big[A_1+T_1 \Big] (x,1-x-x_3,x_3) \nnb \\
\ek 4 e_{q_2} \int_0^{\bar{x}} \,dx_1 \Big[A_1-T_1 \Big] (x_1,x,1-x_1-x) \nnb \\
\ek 8 e_{q_3} \int_0^{\bar{x}}dx_1\, T_1 (x_1,1-x_1-x,x)
\nnb\\ \nnb \\
%
\left( \rho_6\right)_2^{\Sigma^{\ast 0}} (x)\es
16 e_{q_1} m_{\cal O}^2 x \Big[4 m_{\cal O} (1-x) x (\, \check{\!\check{B}}_8 - 
\,\check{\!\check{C}}_6 ) - m_{q_1} \,\check{\!\check{B}}_6 \Big] (x) \nnb \\
\ar 64 e_{q_2} m_{\cal O}^3 (1-x) x^2 \Big[\widetilde{\!\widetilde{B}}_8
- \;\widetilde{\!\widetilde{C}}_6 \Big] (x) \nnb \\
\ar 16 e_{q_3} m_{\cal O}^2 x \Big[ 4 m_{\cal O} (1-x) x \,(\,\widehat{\!\widehat{B}}_8   
- \, \widehat{\!\widehat{C}}_6 ) +                              
m_{q_3} \widehat{\!\widehat{B}}_6 \Big] (x) \nnb \\ \nnb \\
%
\left( \rho_4\right)_2^{\Sigma^{\ast 0}} (x)\es
4 e_{q_1} m_{\cal O} x \Big[2 (1-2 x) (\check{B}_4 - \check{C}_2) - 2 \check{B}_2
+ \check{D}_2 \Big] (x) \nnb \\
\ek 4 e_{q_2} m_{\cal O} x \Big[ 2 \widetilde{B}_2 - 2 (1-2 x)  (\widetilde{B}_4 - \widetilde{C}_2)
+ \widetilde{D}_2 \Big] (x) \nnb \\
\ar 16 e_{q_3} m_{\cal O} x \Big[ (1-x) \widehat{B}_4 + x \widehat{C}_2 \Big] (x) \nnb \\ \nnb \\
%
\left( \rho_6\right)_3^{\Sigma^{\ast 0}} (x)\es
- 64 e_{q_1} m_{\cal O}^3 (1-x)^2 x \Big[\, \check{\!\check{B}}_8  -
\, \check{\!\check{C}}_6\Big] (x) \nnb \\
\ek 64 e_{q_2} m_{\cal O}^3 (1-x)^2 x \Big[ \, \widetilde{\!\widetilde{B}}_8 - \,
\widetilde{\!\widetilde{C}}_6 \Big] (x) \nnb \\
\ek 64 e_{q_3} m_{\cal O}^3 (1-x)^2 x \Big[ \, \widehat{\!\widehat{B}}_8 -
\, \widehat{\!\widehat{C}}_6 \Big] (x)
\nnb \\ \nnb \\
%
\left( \rho_4\right)_3^{\Sigma^{\ast 0}} (x)\es
8 e_{q_1} m_{\cal O} (1-x) \Big[ \check{B}_2 - (1-2 x) \check{B}_4 -  2 x
\check{C}_2 \Big](x) \nnb \\
\ar 8 e_{q_2} m_{\cal O} (1-x) \Big[ \widetilde{B}_2 - (1-2 x) \widetilde{B}_4 - 2 x 
\widetilde{C}_2  \Big] (x) \nnb \\
\ar 8 e_{q_3} m_{\cal O} (1-x) \Big[ \widehat{B}_2 - (1-2 x) \widehat{B}_4 -
2 x \widehat{C}_2 \Big] (x)~, \nnb
\eea
where $q_1=u$, $q_2=d$, and $q_3=s$, respectively.

In the above expressions for $\rho_i$ and $\rho_i^{'}$,
the functions ${\cal F}(x_i)$ are defined in the following way:

\bea
\label{nolabel}
\check{\cal F}(x_1) \es \int_1^{x_1}\!\!dx_1^{'}\int_0^{1- x^{'}_{1}}\!\!dx_3\,
{\cal F}(x_1^{'},1-x_1^{'}-x_3,x_3)~, \nnb \\
\check{\!\!\!\;\check{\cal F}}(x_1) \es 
\int_1^{x_1}\!\!dx_1^{'}\int_1^{x^{'}_{1}}\!\!dx_1^{''}
\int_0^{1- x^{''}_{1}}\!\!dx_3\,
{\cal F}(x_1^{''},1-x_1^{''}-x_3,x_3)~, \nnb \\
\widetilde{\cal F}(x_2) \es \int_1^{x_2}\!\!dx_2^{'}\int_0^{1- x^{'}_{2}}\!\!dx_1\,
{\cal F}(x_1,x_2^{'},1-x_1-x_2^{'})~, \nnb \\
\widetilde{\!\widetilde{\cal F}}(x_2) \es 
\int_1^{x_2}\!\!dx_2^{'}\int_1^{x^{'}_{2}}\!\!dx_2^{''}
\int_0^{1- x^{''}_{2}}\!\!dx_1\,
{\cal F}(x_1,x_2^{''},1-x_1-x_2^{''})~, \nnb \\
\widehat{\cal F}(x_3) \es \int_1^{x_3}\!\!dx_3^{'}\int_0^{1- x^{'}_{3}}\!\!dx_1\,
{\cal F}(x_1,1-x_1-x_3^{'},x_3^{'})~, \nnb \\
\widehat{\!\widehat{\cal F}}(x_3) \es 
\int_1^{x_3}\!\!dx_3^{'}\int_1^{x^{'}_{3}}\!\!dx_3^{''}
\int_0^{1- x^{''}_{3}}\!\!dx_1\,
{\cal F}(x_1,1-x_1-x_3^{''},x_3^{''})~.\nnb
\eea

Definitions of the functions $B_i$, $C_i$, $D_i$, $E_1$ and $H_1$
that appear in the expressions for $\rho_i(x)$ are given as follows:

\bea
\label{nolabel}
B_2 \es T_1+T_2-2 T_3~, \nnb \\
B_4 \es T_1-T_2-2 T_7~, \nnb \\
B_5 \es - T_1+T_5+2 T_8~, \nnb \\
B_6 \es 2 T_1-2 T_3-2 T_4+2 T_5+2 T_7+2 T_8~, \nnb \\
B_7 \es T_7-T_8~, \nnb \\
B_8 \es  -T_1+T_2+T_5-T_6+2 T_7+2T_8~, \nnb \\
C_2 \es V_1-V_2-V_3~, \nnb \\
C_4 \es -2V_1+V_3+V_4+2V_5~, \nnb \\
C_5 \es V_4-V_3~, \nnb \\
C_6 \es -V_1+V_2+V_3+V_4+V_5-V_6~, \nnb \\
D_2 \es -A_1+A_2-A_3~, \nnb \\
D_4 \es -2A_1-A_3-A_4+2A_5~, \nnb \\
D_5 \es A_3-A_4~, \nnb \\
D_6 \es A_1-A_2+A_3+A_4-A_5+A_6~, \nnb \\
E_1 \es S_1-S_2~, \nnb \\
H_1 \es P_2-P_1~. \nnb
\eea

\newpage

\section*{Figure captions}
{\bf Fig. (1)} The dependence of the magnetic dipole form factor 
$G_M(Q^2)$ of the $\gamma^\ast \Sigma^+ \to \Sigma^{\ast +}$ 
transition on $Q^2$ at $s_0=4.0~GeV^2$ and 
at several different fixed values of the Borel mass parameter $M^2$.\\ \\
{\bf Fig. (2)} The same as in Fig. (1), but for the electric quadrupole
$G_E(Q^2)$ form factor.\\ \\
{\bf Fig. (3)} The same as in Fig. (1), but for the Coulomb
quadrupole $G_C(Q^2)$ form factor.\\ \\
{\bf Fig. (4)} The same as in Fig. (1), but for the $\gamma^\ast \Lambda \to
\Sigma^{\ast 0}$ transition.\\ \\
{\bf Fig. (5)} The same as in Fig. (1), but for the $\gamma^\ast \Sigma^0
\to \Sigma^{\ast 0}$ transition.\\ \\
{\bf Fig. (6)} The same as in Fig. (1), but for the $\gamma^\ast \Sigma^-
\to \Sigma^{\ast -}$ transition.\\ \\
{\bf Fig. (7)} The same as in Fig. (1), but for the $\gamma^\ast
\Xi^- \to \Xi^{\ast -}$ transition.\\ \\
{\bf Fig. (8)} The same as in Fig. (1), but for the $\gamma^\ast \Xi^0 \to
\Xi^{\ast 0}$ transition.\\ \\
{\bf Fig. (9)} The dependence of the the ratio $R_{EM}$ on $Q^2$ for the
octet to decuplet electromagnetic transitions, at $s_0=4.0~GeV^2$ and
$M^2=2~GeV^2$.\\ \\
{\bf Fig. (10)} The same as in Fig. (9), but for the ratio $R_{SM}$.

\newpage

\begin{figure}
\vskip 3. cm
    \includegraphics{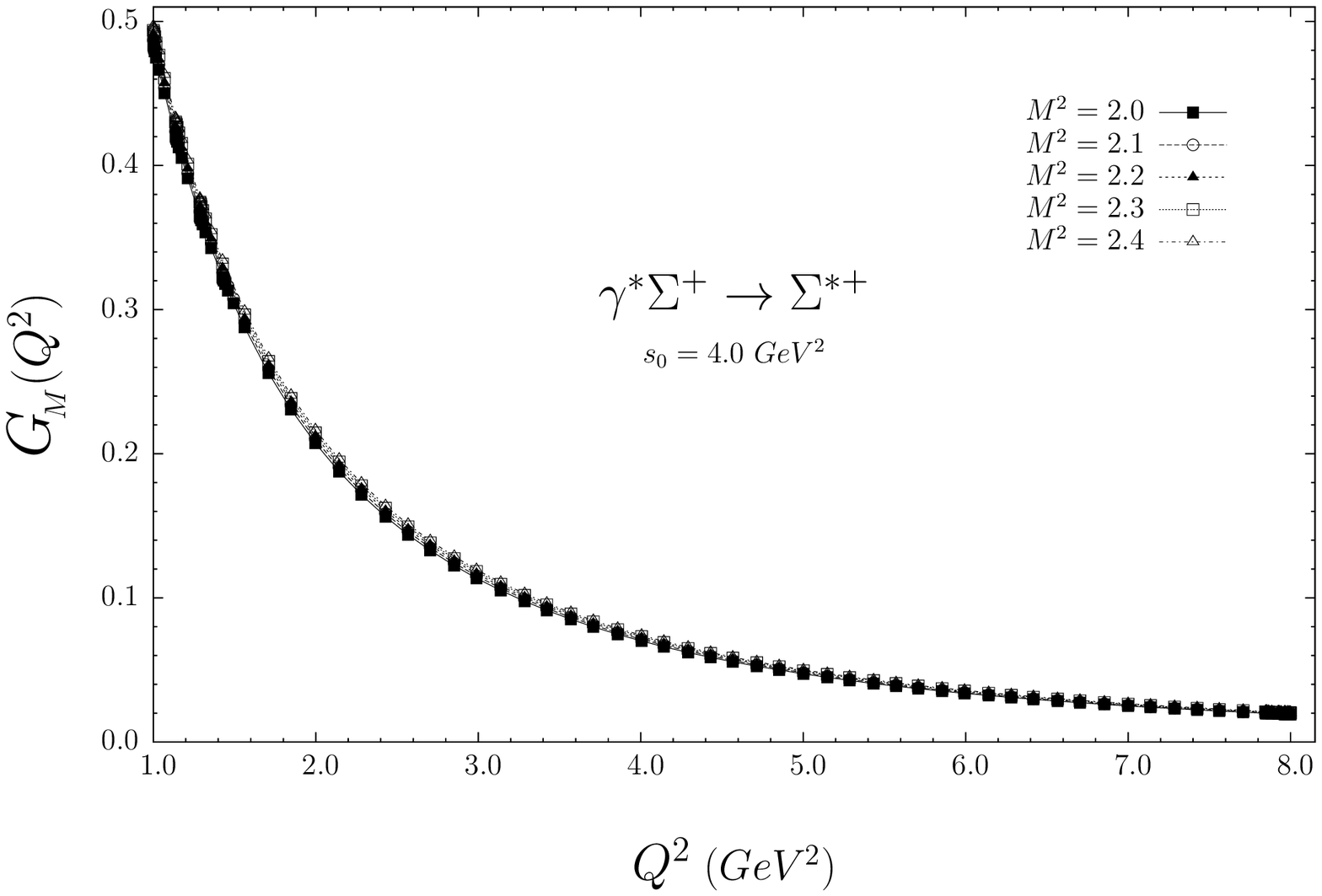}
\vskip 7.0cm
\caption{}
\end{figure}

\begin{figure}
\vskip 3. cm
    \includegraphics{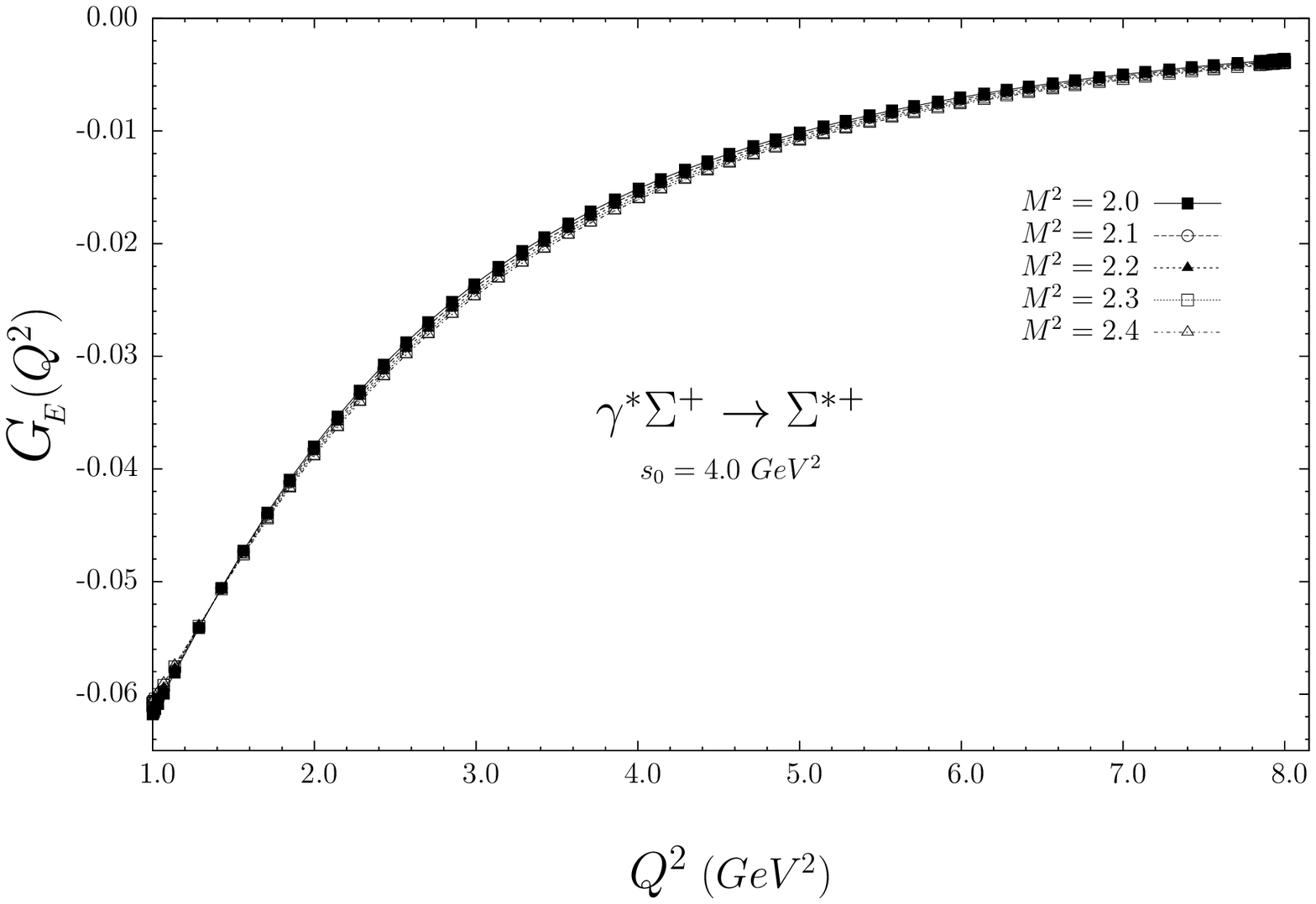}
\vskip 7.0cm
\caption{}
\end{figure}

\begin{figure}
\vskip 3. cm
    \includegraphics{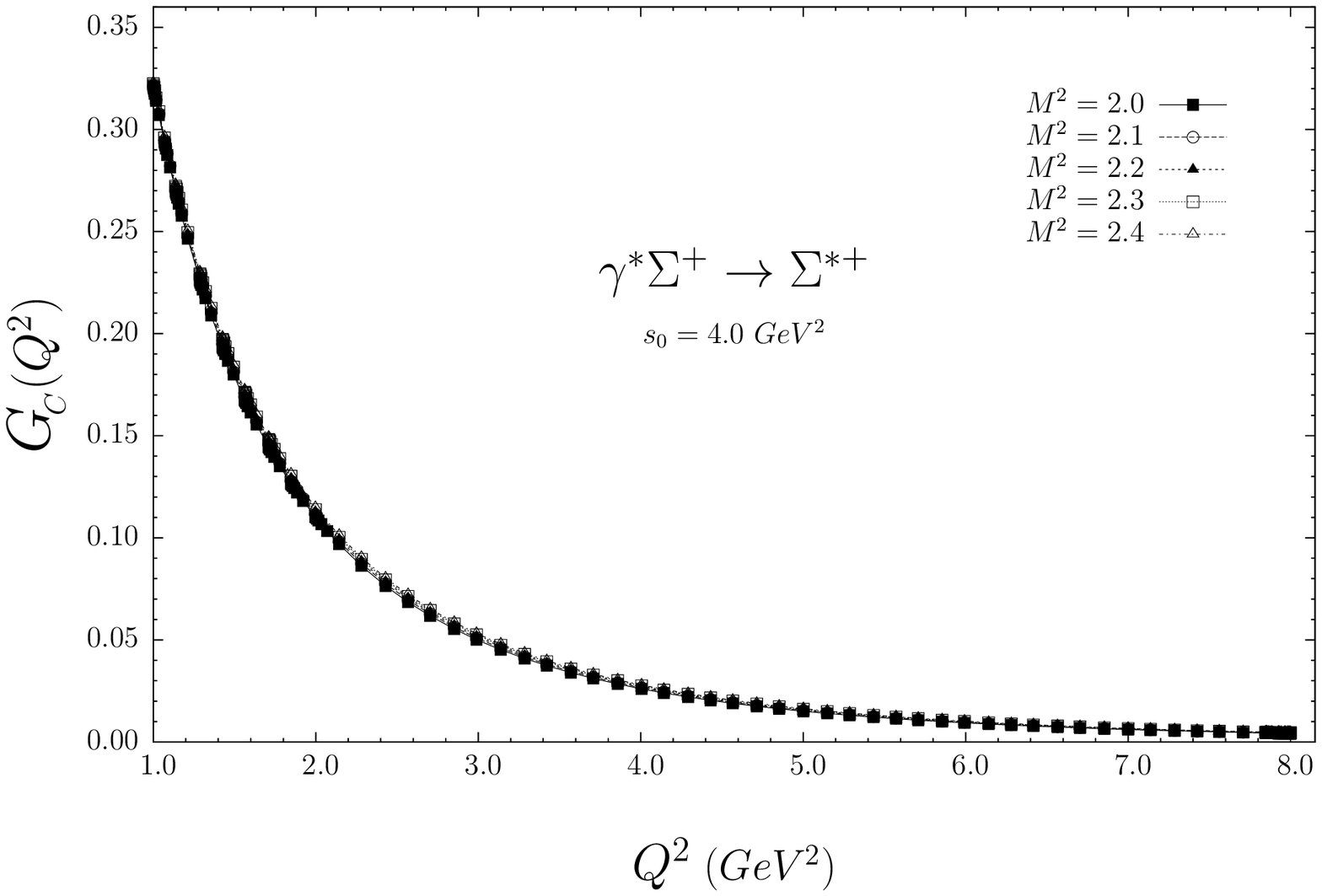}
\vskip 7.0cm
\caption{}
\end{figure}

\begin{figure}
\vskip 3. cm
    \includegraphics{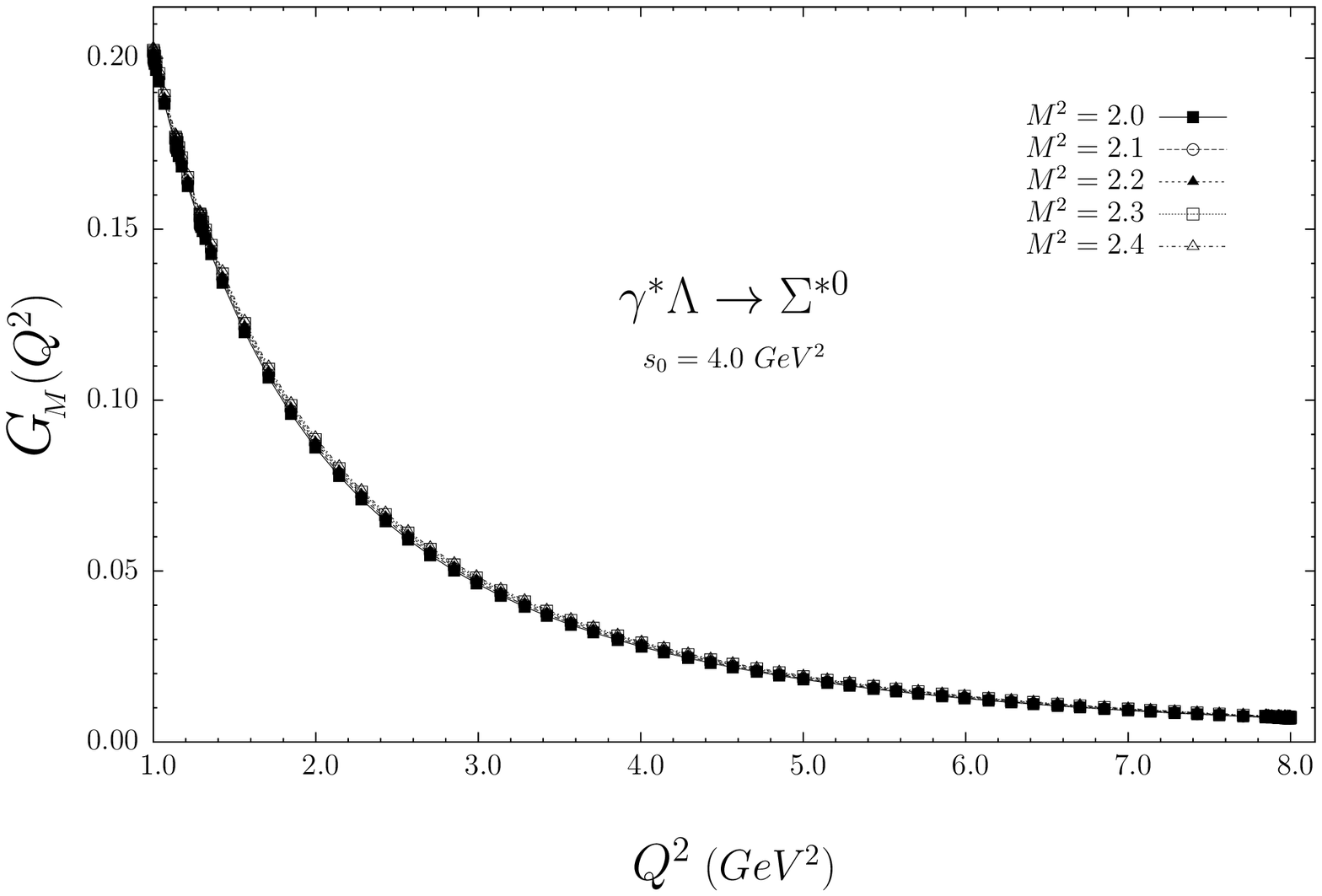}
\vskip 7.0cm
\caption{}
\end{figure}

\begin{figure}
\vskip 3. cm
    \includegraphics{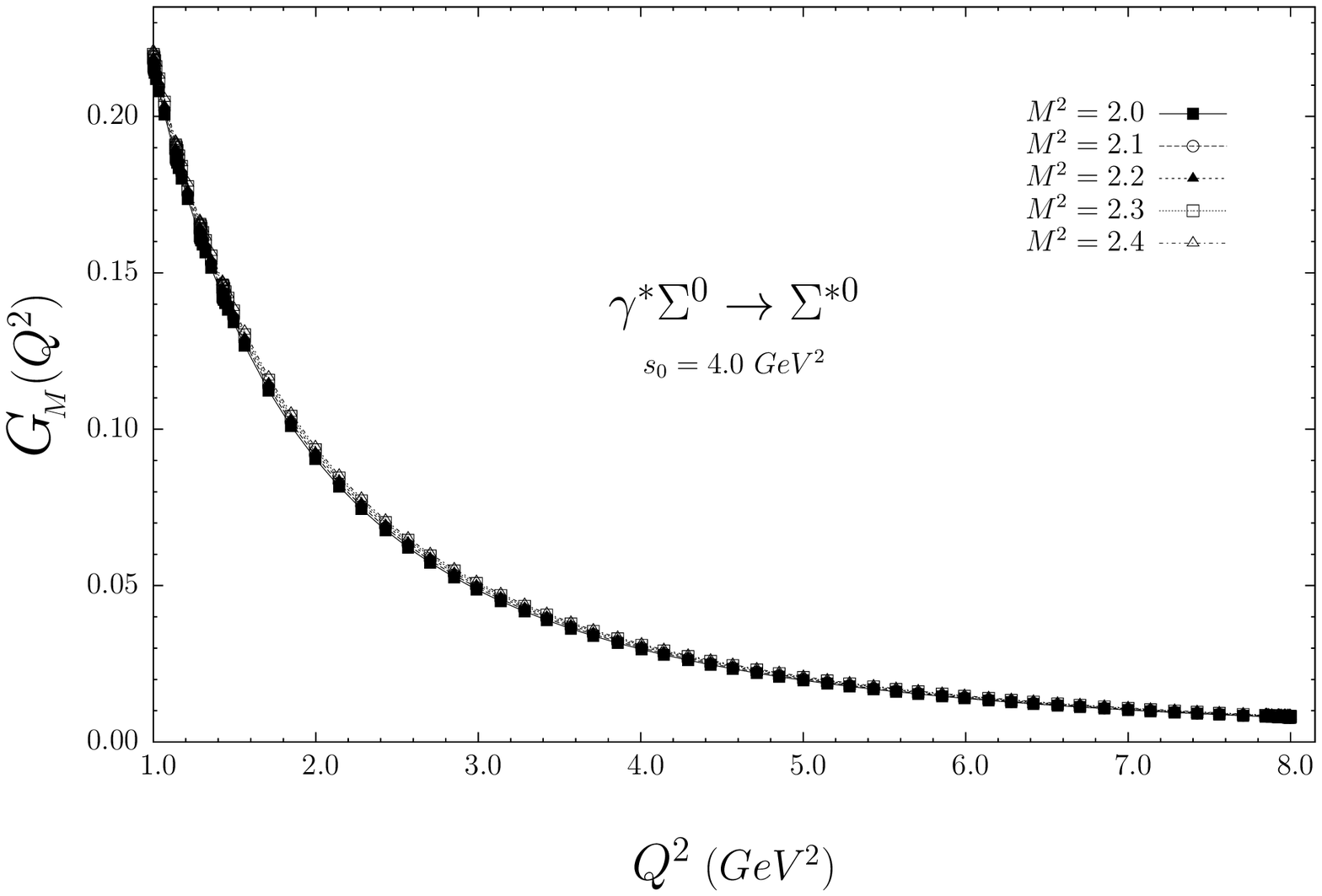}
\vskip 7.0cm
\caption{}
\end{figure}

\begin{figure}
\vskip 3. cm
    \includegraphics{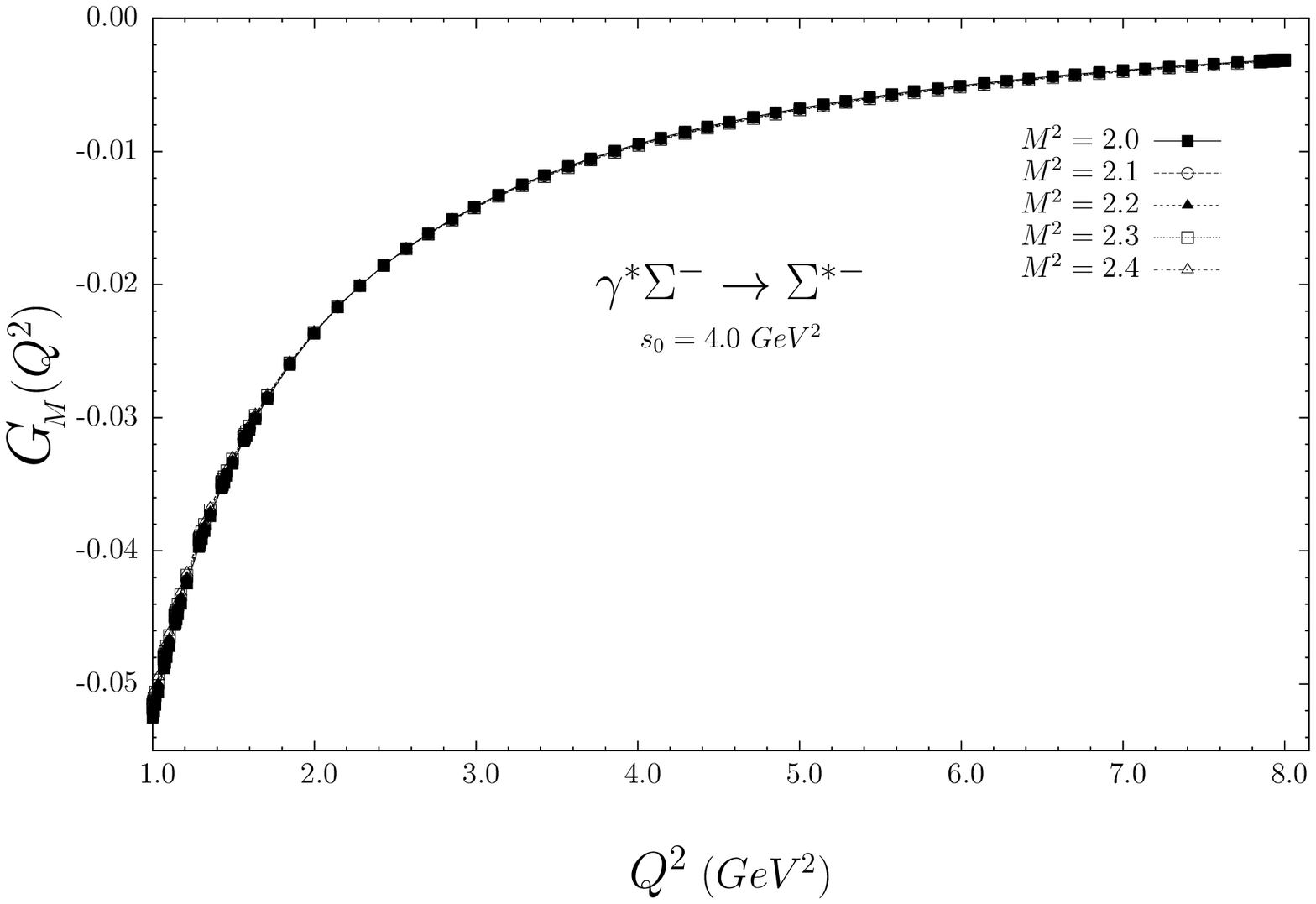}
\vskip 7.0cm
\caption{}
\end{figure}

\begin{figure}
\vskip 3. cm
    \includegraphics{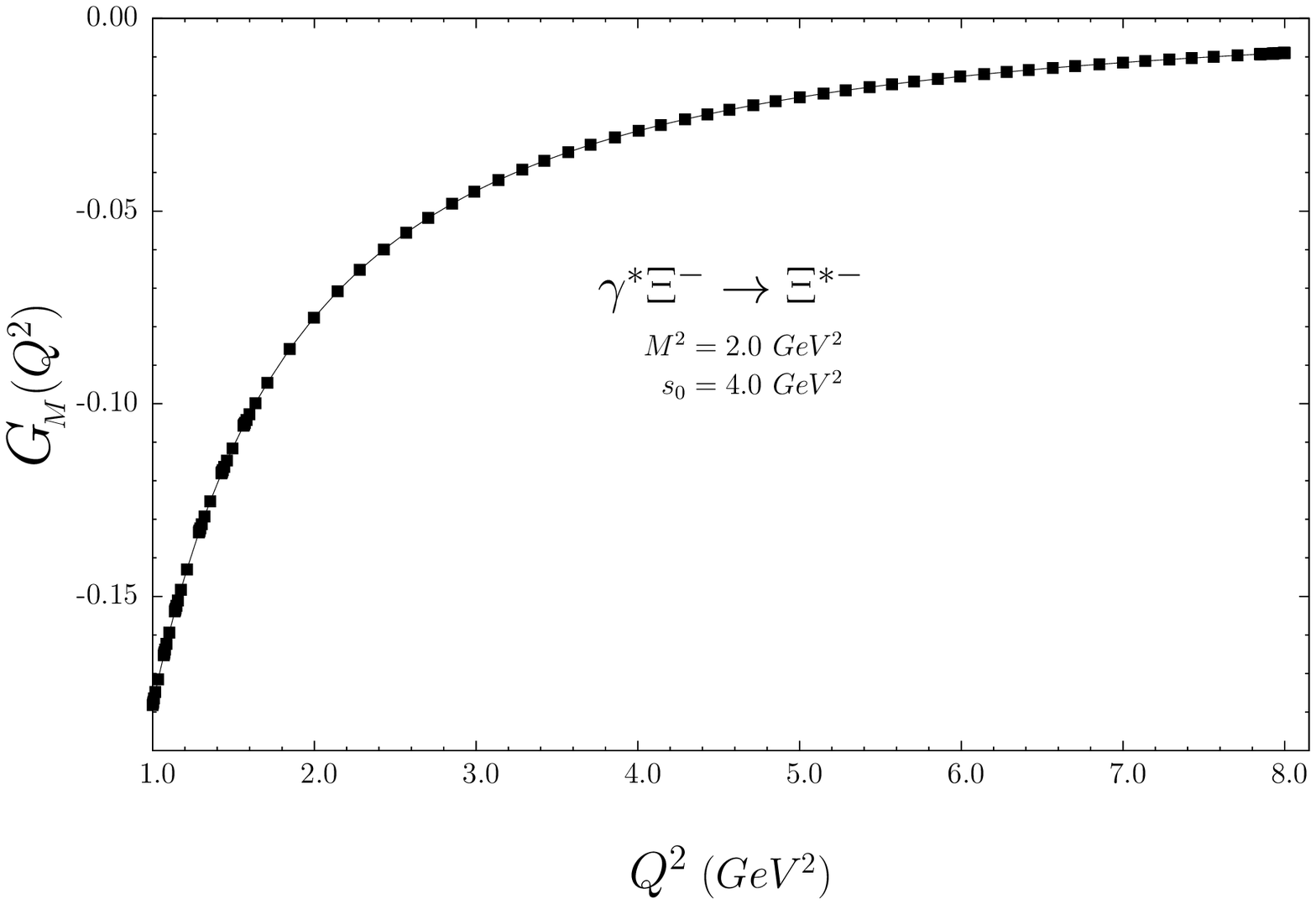}
\vskip 7.0cm
\caption{}
\end{figure}

\begin{figure}
\vskip 3. cm
    \includegraphics{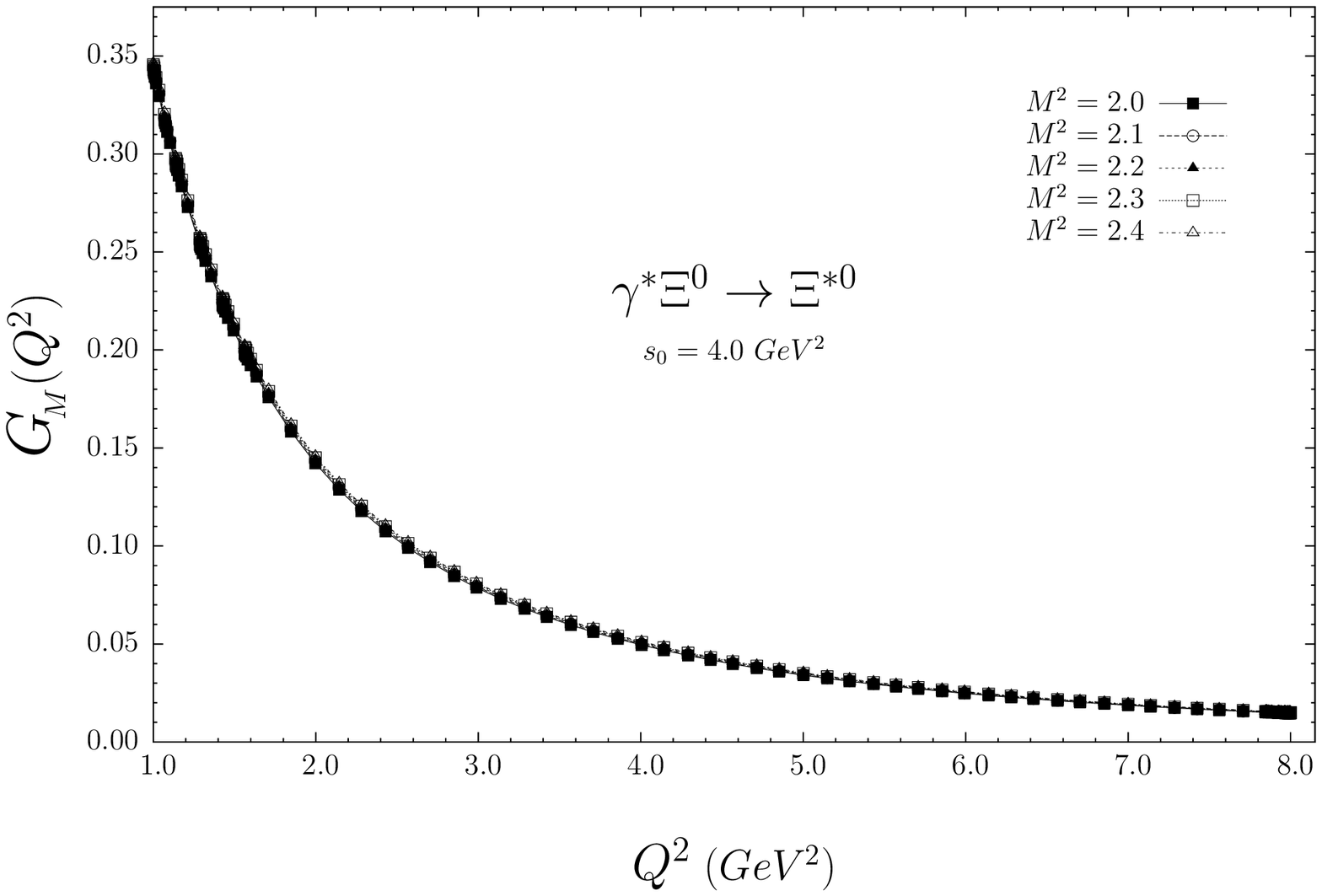}
\vskip 7.0cm
\caption{}
\end{figure}

\begin{figure}
\vskip 3. cm
    \includegraphics{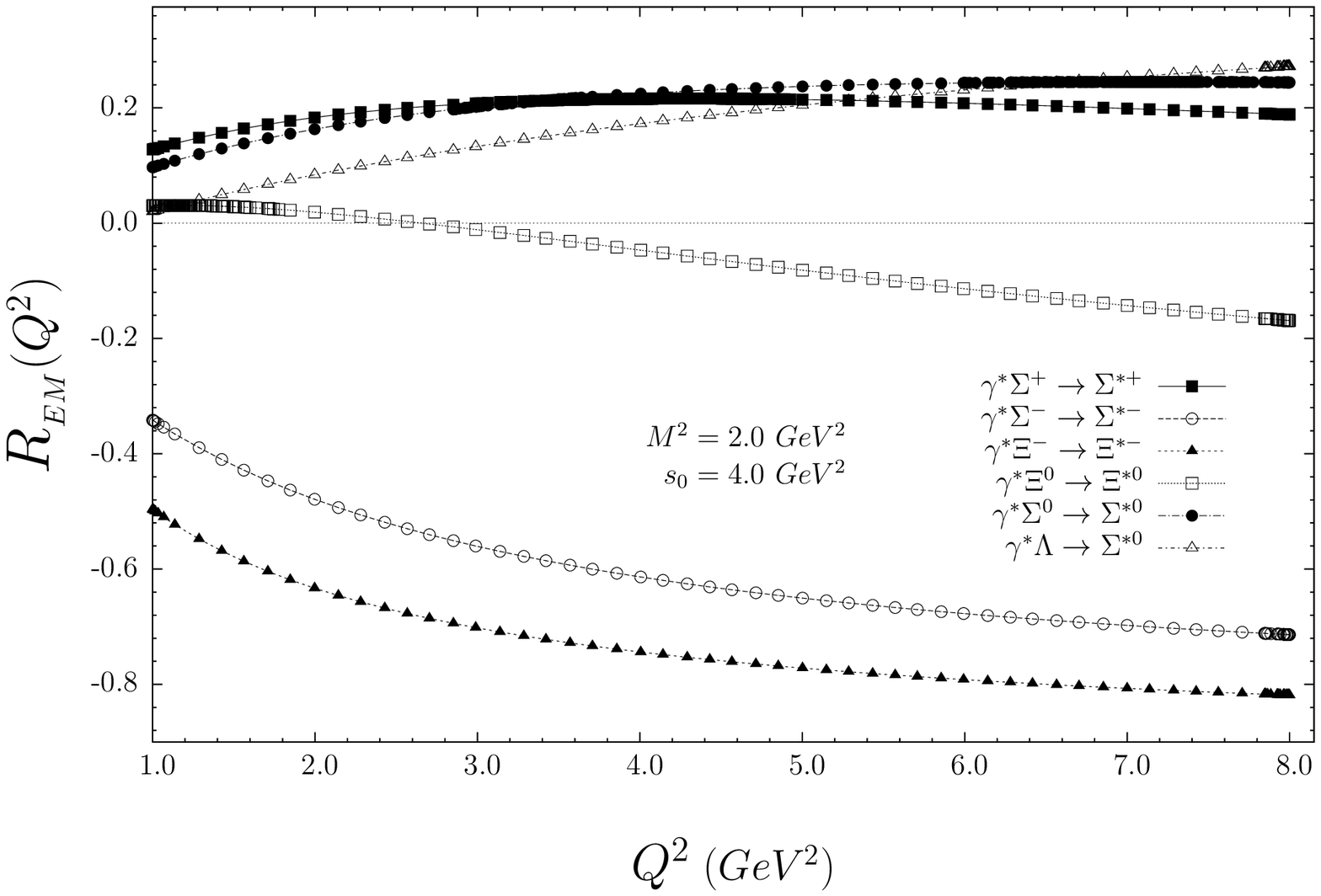}
\vskip 7.0cm
\caption{}
\end{figure}

\begin{figure}
\vskip 3. cm
    \includegraphics{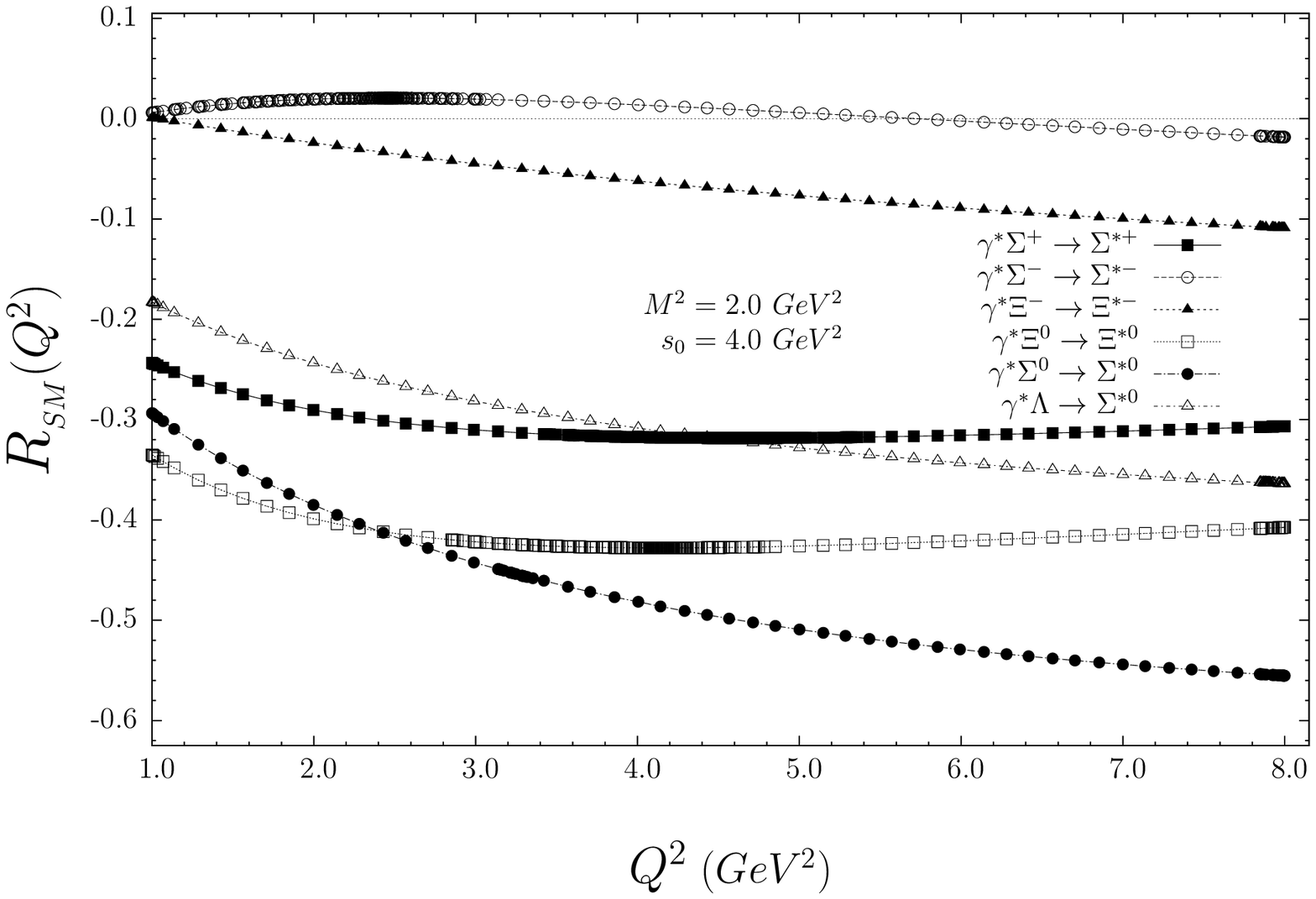}
\vskip 7.0cm
\caption{}
\end{figure}

\end{document}